\documentclass{article}
\usepackage{spconf,amsmath,graphicx,xcolor}

\title{Multi-Task Pseudo-Label Learning for Non-Intrusive Speech Quality Assessment Model}
\name{Ryandhimas E. Zezario$^1$$^2$, Bo-Ren Brian Bai$^3$, Chiou-Shann Fuh$^1$, Hsin-Min Wang$^2$, Yu Tsao$^2$}
\address{$^1$National Taiwan University $^2$Academia Sinica $^3$Fortemedia}

\begin{document}
\maketitle
\begin{abstract}
This study proposes a multi-task pseudo-label learning (MPL)-based non-intrusive speech quality assessment model called MTQ-Net. MPL consists of two stages: obtaining pseudo-label scores from a pretrained model and performing multi-task learning. The 3QUEST metrics, namely Speech-MOS (S-MOS), Noise-MOS (N-MOS), and General-MOS (G-MOS), are the assessment targets. The pretrained MOSA-Net model is utilized to estimate three pseudo labels: perceptual evaluation of speech quality (PESQ), short-time objective intelligibility (STOI), and speech distortion index (SDI). Multi-task learning is then employed to train MTQ-Net by combining a supervised loss (derived from the difference between the estimated score and the ground-truth label) and a semi-supervised loss (derived from the difference between the estimated score and the pseudo label), where the Huber loss is employed as the loss function. Experimental results first demonstrate the advantages of MPL compared to training a model from scratch and using a direct knowledge transfer mechanism. Second, the benefit of the Huber loss for improving the predictive ability of MTQ-Net is verified. Finally, the MTQ-Net with the MPL approach exhibits higher overall predictive power compared to other SSL-based speech assessment models. 

\end{abstract}
\section{Introduction}
\label{sec:intro}

Speech assessment metrics are important quantitative evaluation indicators for speech-related applications, such as speech synthesis \cite{huang2022voicemos}, speech enhancement (SE) \cite{loizou2007speech}, hearing aids \cite {barker20221st}, and telecommunications \cite {yi22b_interspeech}. With the emergence of deep learning and the availability of training samples, researchers have begun to employ deep learning models to deploy speech assessment metrics in different tasks, e.g., voice conversion \cite{mbnet_mos, ssl-mos}, speech enhancement \cite{mosa-net, TMINT-QI}, hearing aids \cite{barker22_interspeech, zezario2022mbi}, and telecommunications \cite{yi22b_interspeech}. To achieve more accurate automatic assessment, several strategies have also been explored, e.g., reducing the bias per listener \cite{mbnet_mos}, incorporating a self-supervised learning model \cite{ssl-mos, mosa-net}, and performing ensemble learning \cite{yang22o_interspeech, saeki22c_interspeech}. Despite significant improvements in performance, achieving satisfactory generalization with a limited number of training samples remains a challenge. The main objective of this study is to explore how information from an established deep learning-based speech assessment model trained on a larger training set can be leveraged to improve prediction performance on a target assessment task with limited training data. In our previous work \cite{mosa-net}, we proposed  MOSA-Net, a multi-objective speech assessment model that uses cross-domain features (spectral and temporal features) and latent representations from an SSL model \cite{hubert} to simultaneously predict objective quality, intelligibility, and distortion scores. We also found that, through knowledge transferring, MOSA-Net trained on objective assessment metrics can be successfully adapted to predict subjective quality and intelligibility scores. 

This study aims to further expand the applicability of MOSA-Net as a teacher model by introducing the multi-task pseudo-label learning (MPL) approach. MPL consists of two stages: obtaining pseudo-label scores and performing multi-task learning. MPL uses both supervised loss and semi-supervised loss to train the target model, where the supervised loss estimates the difference between the predicted score and the ground-truth label score, and the semi-supervised loss estimates the difference between the predicted score and the pseudo-label score. Additionally, the Huber loss \cite{huber}, which combines mean absolute error (MAE) and mean square error (MSE), is employed as the loss function. The pseudo-label scores used in MPL include perceptual evaluation of speech quality (PESQ), short-time objective intelligibility (STOI), and speech distortion index (SDI) \cite{sdi} obtained from the pretrained MOSA-Net model. In this study, we investigate the application of MPL in transferring knowledge of MOSA-Net to deploy a multi-target speech quality assessment network called MTQ-Net. The primary objective of MTQ-Net is to simultaneously predict three 3QUEST metrics \cite{3quests}, namely Speech-MOS (S-MOS), Noise-MOS (N-MOS), and General-MOS (G-MOS) scores, which are widely used in telecommunications. 
There are paid costs for using the 3QUEST tool. Additionally, it is impossible to estimate the 3QUEST scores of any speech utterance without a corresponding clean speech. To overcome these limitations, it is highly desirable to train a neural network that can estimate the 3QUEST metrics from a single utterance.
Our experimental results demonstrate the advantages of MPL over knowledge transfer and training from scratch approaches, allowing the deployed MTQ-Net model to achieve better prediction capabilities. Furthermore, utilizing the Huber loss can yield higher prediction performance compared to MAE and MSE alone. Finally, MTQ-Net with the MPL approach demonstrates improved overall prediction performance compared to other SSL-based speech assessment models \cite{ssl-mos}.

The remainder of this paper is organized as follows. Section II presents the proposed MPL mechanism and its use in MTQ-Net. Section III describes the experimental setup and results. Finally, Section IV presents the conclusions and future work. 

\section{Multi-Task Pseudo-Label Learning}
\label{sec:MTQ}

\begin{figure}[t]
\centering
\includegraphics[width=8cm]{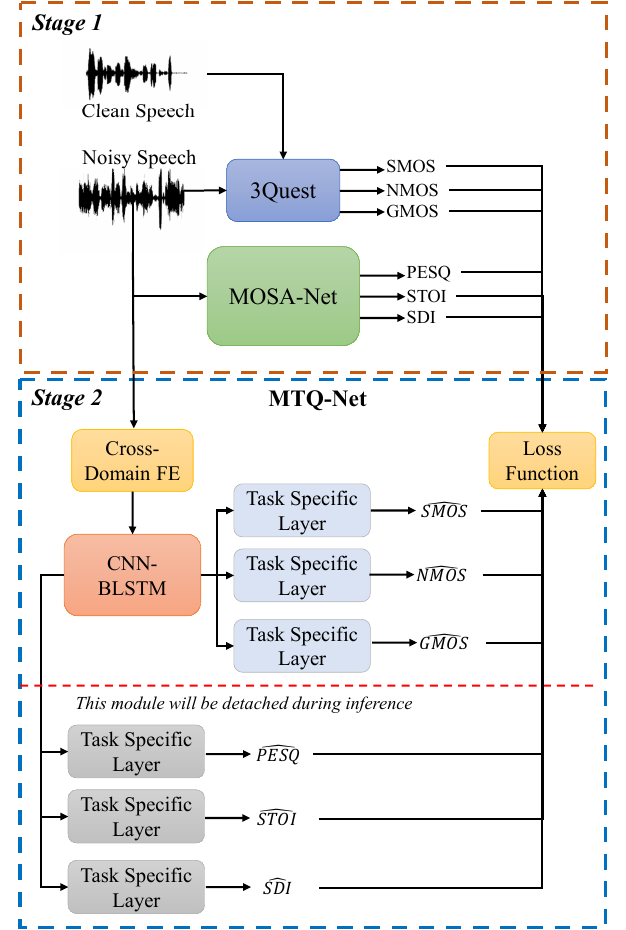} 
\caption{Overall framework of MPL for training the MTQ-Net model.} 
\label{fig:mtq}
\end{figure}

In this section, we introduce the overall framework of the MPL approach. As shown in Fig.~\ref{fig:mtq}, MPL consists of two distinct stages: obtaining pseudo-label scores and performing multi-task learning. The primary ground-truth labels are obtained using the 3QUEST tool, where the calculation of the S-MOS, N-MOS, and G-MOS scores of a speech utterance requires its corresponding clean speech utterance. The pretrained MOSA-Net model is adopted to obtain the pseudo labels, including PESQ, STOI, and SDI scores.

In the second stage, multi-task learning is performed to train the MTQ-Net model to predict 3QUEST scores. As shown in Fig.~\ref{fig:mtq}, during the training phase, speech waveforms are processed by a cross-domain feature extraction module. This module generates three types of acoustic features: power spectral features from the short-time Fourier transform (STFT), learnable filter banks (LFB) from the Sinc convolution layer \cite{sincnet}, and SSL embeddings from a self-supervised learning (SSL) model \cite{chen2021wavlm}. These three acoustic features are then processed by the CNN-BLSTM module as in MOSA-Net \cite{mosa-net}. In the final step, the output of the CNN-BLSTM module is processed by six different task-specific layers aimed at estimating S-MOS, N-MOS, G-MOS, PESQ, STOI, and SDI scores, respectively. Specifically, each task-specific layer consists of an attention layer, a fully connected layer, and a global average pooling layer. It is worth noting that the three task-specific layers for estimating pseudo-label scores will be detached during inference. The purpose of adding these three additional task-specific layers is to improve the generalization ability of the encoder layer during the training phase. The objective function for training the MTQ-Net model is defined as follows:

\graphicspath{ {./images/} }

\begin{equation}
\label{eq:loss}
    \begin{array}{c}
    O =  L_{superv} + L_{semi} \\
    L_{superv} =  L_{SMOS} + L_{NMOS} +  L_{GMOS}  \\  
    L_{semi} =  L_{PESQ} + L_{STOI} +  L_{SDI}        
    \\
    \end{array} 
\end{equation}
The overall loss $O$ consists of a supervised loss $L_{superv}$ calculated based on the ground-truth scores of the primary (i.e., target) task and a semi-supervised loss $L_{semi}$ calculated based on the pseudo-label scores of the auxiliary task. For each assessment metric (e.g., S-MOS), the loss (e.g., $L_{SMOS}$) is calculated by adding utterance-level loss $L_{{utt}}$ and frame-level loss $L_{{fr}}$, following \cite{mosa-net}. $L_{{utt}}$ and $L_{{fr}}$ are calculated as follows: 

\begin{equation}
\label{eq:loss3}
\small
L_{{utt}}=\left\{ 
  \begin{array}{ c l }
  \frac{1}{U}\sum\limits_{u=1}^U \frac{1}{2}(S_u-\hat{S}_u)^2& \quad  |S_u-\hat{S}_u| \leq \delta \\
  
  \frac{1}{U}\sum\limits_{u=1}^U \delta(|S_u-\hat{S}_u|-\frac{1}{2}\delta) & \quad \textrm{otherwise}
  \end{array}
  \right.  
\end{equation}

\begin{equation}
\small
\label{eq:loss4}
L_{{fr}}=\left\{ 
  \begin{array}{ c l }
  \frac{1}{U}\sum\limits_{u=1}^U \frac{\alpha_S}{F_u}\sum\limits_{f=1}^{F_u} \frac{1}{2}(S_u-\hat{s}_f)^2 & \quad |S_u-\hat{s}_f| \leq \delta \\
  \frac{1}{U}\sum\limits_{u=1}^U \frac{\alpha_S}{F_u}\sum\limits_{f=1}^{F_u} \delta(|S_u-\hat{s}_f)|-\frac{1}{2}\delta) & \quad \textrm{otherwise}
  \end{array}
  \right. \\
\end{equation}
$S_u, \hat{S}_u, \hat{s}_f $ are the ground-truth (or pseudo-label) score, predicted utterance-level score, and predicted frame-level score, respectively. The parameter $\delta$ is a hyperparameter that determines whether the Huber loss uses MAE or MSE. $U$ denotes the total number of training utterances; ${F}_u$ denotes the number of frames in the $u$-th training utterance; $\alpha_S$ is the weight between utterance-level and frame-level losses.

\section{Experiments}
\subsection{Experimental Setup}
We evaluated the proposed MTQ-Net model on the Taiwan Mandarin Hearing In Noise test - Quality $\&$ Intelligibility (TMHINT-QI) dataset \cite{TMINT-QI}. The dataset includes clean, noisy, and enhanced speech utterances from five different SE systems, including Karhunen-Loeve transform (KLT) \cite{klt}, minimum-mean squared error (MMSE) \cite{mmse}, fully convolutional network (FCN) \cite{FCN}, deep denoising autoencoder (DDAE) \cite{DDAE}, and transformer-based SE \cite{Trans}. TMHINT-QI provided a diverse set of metrics, including both objective and subjective measures. In this study, we specifically concentrated on additional metrics, not originally part of TMHINT-QI, which are the 3QUEST scores, consisting of S-MOS, N-MOS, and G-MOS scores. We prepared training labels for MTQ-Net by calculating 3QUEST scores based on the noisy-clean or enhanced-clean paired utterances from TMHINT-QI. It's important to emphasize that we operated under the assumption that the other metrics offered by TMHINT-QI were inaccessible for our analysis.

The training set contained 11,000 utterances with corresponding S-MOS, N-MOS, and G-MOS scores as ground-truth labels. Specifically, from 11,000 speech samples, we allocated 90\% for training and 10\% for validation. The test set contained 2,500 utterances with corresponding ground-truth labels. It is worth noting that there is no overlap in training and test utterances. We used three evaluation metrics, namely MSE, linear correlation coefficient (LCC), and Spearman's rank correlation coefficient (SRCC) \cite{srcc} to evaluate the prediction output of all compared models. The smaller the difference between the predicted score and the ground-truth score, the smaller the MSE value; therefore, a lower MSE value indicates better performance. The LCC and SRCC values respectively represent the numerical or ranked correlation between the predicted scores and the ground-truth scores; so the higher the value, the higher the correlation and the better the performance.

\subsection{MTQ-Net with Different Training Mechanisms}
In the first experiment, we compared three different training mechanisms for building MTQ-Net. First, we trained MTQ-Net from scratch using three types of ground-truth labels (denoted as ``From Scratch''). Second, we performed simple knowledge transfer, i.e., initialized MTQ-Net with the weights of the MOSA-Net model trained to predict PESQ, STOI, and SDI, and then trained the model using three types of ground-truth labels (denoted as ``KT''). Third, we adopted the MPL approach to construct MTQ-Net (denoted as ``MPL''). In addition, we also compared the embeddings of two different SSL models, namely HuBERT \cite{hubert} and WavLM \cite{chen2021wavlm}. All models were trained using the MSE loss. From Table \ref{tab:diff_train}, we first notice that performing knowledge transfer helps MTQ-Net achieve better prediction performance across all evaluation metrics (MTQ-Net (HuBERT)-KT vs MTQ-Net (HuBERT)-From Scratch). Second, WavLM embeddings are slightly more effective than HuBERT embeddings in most cases, although not always (MTQ-Net (WavLM)-KT vs MTQ-Net (HuBERT)-KT). Third, MPL, which considers three additional pseudo labels, is indeed more effective than simple knowledge transfer (MTQ-Net (WavLM)-MPL vs MTQ-Net (WavLM)-KT). Fourth, fine-tuning the SSL model during MTQ-Net training can yield better performance in most cases (MTQ-Net (FT-WavLM)-MPL vs MTQ-Net (WavLM)-MPL).

 \begin{table}[t]
\caption{LCC, SRCC, and MSE results of MTQ-Net using different training mechanisms.}
\label{tab:diff_train}
\footnotesize
\begin{center}
\setlength\tabcolsep{3pt}
 \begin{tabular}{c||c||c||c||c} 
 \hline
 \hline
 \textbf{Systems} &\textbf{Method} &\textbf{LCC} & \textbf{SRCC} & \textbf{MSE}  \\ [0.5ex] \cline{2-4}
 \hline\hline
  \multicolumn{5}{c} {S-MOS Score Prediction} \\
 \hline
MTQ-Net (HuBERT)&From Scratch&0.891&0.884&0.055\\ \hline
MTQ-Net (HuBERT)&KT&0.894&0.892&0.055\\ \hline
MTQ-Net (WavLM)&KT&0.899&0.887&0.048\\ \hline
MTQ-Net (WavLM)&MPL&0.902&0.895&0.046\\ \hline
MTQ-Net (FT-WavLM)&MPL&\textbf{0.913}&\textbf{0.908}&\textbf{0.045}\\ \hline
 \hline
  \multicolumn{5}{c} {N-MOS Score Prediction} \\
 \hline
MTQ-Net (HuBERT)&From Scratch&0.734&0.775&0.107\\ \hline
MTQ-Net (HuBERT)&KT&\textbf{0.745}&0.787&\textbf{0.102}\\ \hline
MTQ-Net (WavLM)&KT&0.739&0.786&0.115\\ \hline
MTQ-Net (WavLM)&MPL&0.737&\textbf{0.788}&0.123\\ \hline
MTQ-Net (FT-WavLM)&MPL&0.714&0.775&0.115\\ \hline
 \hline
  \multicolumn{5}{c} {G-MOS Score Prediction} \\
 \hline
MTQ-Net (HuBERT)&From Scratch&0.851&0.850&0.047\\ \hline
MTQ-Net (HuBERT)&KT&0.860&0.863&0.045\\ \hline
MTQ-Net (WavLM)&KT&0.868&0.865&0.044\\ \hline
MTQ-Net (WavLM)&MPL&0.876&\textbf{0.880}&0.043\\ \hline
MTQ-Net (FT-WavLM)&MPL&\textbf{0.877}&0.876&\textbf{0.042}\\ \hline
 \hline
\end{tabular}
\end{center}
\end{table}

\begin{table}[t]
\caption{LCC, SRCC, and MSE results of MTQ-Net using different loss functions.}
\footnotesize
\label{tab:diff_loss}
\begin{center}
 \begin{tabular}{c||c||c||c||c} 
 \hline
 \hline
 \textbf{Method} &\textbf{Target} &\textbf{LCC} & \textbf{SRCC} & \textbf{MSE}  \\ [0.5ex] \cline{2-4}
 \hline\hline
  \multicolumn{5}{c} {S-MOS Score Prediction} \\
 \hline
MPL&MSE&\textbf{0.913}&\textbf{0.908}&0.045\\ \hline
MPL&MAE&0.904&0.897&0.074\\ \hline
MPL&Huber&0.912&0.903&\textbf{0.043}\\ \hline
 \hline
  \multicolumn{5}{c} {N-MOS Score Prediction} \\
 \hline
MPL&MSE&0.714&0.775&0.115\\ \hline
MPL&MAE&0.712&0.783&0.117\\ \hline
MPL&Huber&\textbf{0.739}&\textbf{0.789}&\textbf{0.108}\\ \hline 
 \hline
  \multicolumn{5}{c} {G-MOS Score Prediction} \\
 \hline
MPL&MSE&0.877&0.876&0.042\\ \hline
MPL&MAE&0.864&0.864&0.047\\ \hline
MPL&Huber&\textbf{0.881}&\textbf{0.882}&\textbf{0.039}\\ \hline
\end{tabular}
\end{center}
\end{table}

\subsection{MTQ-Net with Different Loss Functions}
In the second experiment, we compared three different loss functions used in MPL for training MTQ-Net, namely MAE, MSE, and Huber loss. MAE is known to be robust against outlier data, while MSE performs appropriate calculations by accommodating small errors equally important as large errors. On the other hand, Huber loss combines the advantages of MAE and MSE and selects the most appropriate loss through the parameter $\delta$. In this experiment, $\delta$ was set to 1.0. As shown in Table \ref{tab:diff_loss}, using MSE loss can achieve better overall performance than using MAE loss. Compared with MAE and MSE alone, Huber loss combining MAE and MSE can enable the MTQ-Net model to achieve better overall performance.

\subsection{Huber Loss with Different $\delta$ Values}
\begin{figure}[t]
\centering
\includegraphics[width=8.5cm]{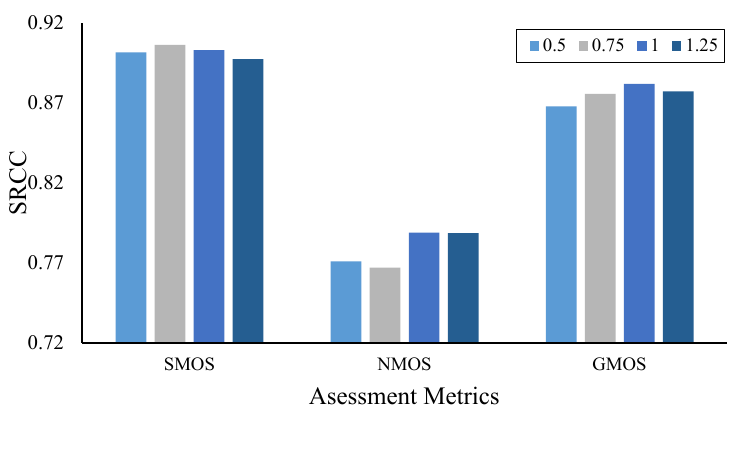} 
\vspace{-0.7cm}
\caption{SRCC results of MTQ-Net trained using Huber loss with different $\delta$ values.} 
\label{fig:hub}
\end{figure}

One of the main important mechanisms of Huber Loss is the flexibility to switch between MAE and MSE through the parameter $\delta$. Specifically, the Huber loss behaves like the MAE if the absolute difference between the label and the predicted score is larger than $\delta$. Otherwise, the Huber loos behaves like the MSE. To study the optimal value of $\delta$, we deployed MTQ-Net using four different $\delta$ values (0.5, 0.75, 1.00, and 1.25). As shown in Fig. \ref{fig:hub}, MTQ-Net trained with different $\delta$ values obtains different SRCC values in predicting 3QUEST scores. For S-MOS prediction, the best performance is obtained when $\delta$ is set to 0.75, while for N-MOS and G-MOS prediction, the best performance is obtained when $\delta$ is set to 1.0. This result shows that setting $\delta$ to 1.0 yields the best overall performance.

\graphicspath{ {./images/} }

\subsection{Comparing MTQ-Net with Other Models}
For a more comprehensive study, we compared MTQ-Net with another SSL-based model (i.e., MOS-SSL \cite{ssl-mos}). MOS-SSL was built on the pretrained wav2vec 2.0 model by mean-pooling its output embeddings and adding a linear output layer to predict MOS scores. The wav2vec 2.0 model was fine-tuned during MOS-SSL training. 
We trained three versions of MOS-SSL to predict S-MOS, N-MOS, and G-MOS, respectively. We chose MOS-SSL as the baseline due to its remarkable performance as a baseline in the VoiceMOS Challenge 2022. 
Since most recent speech assessment models employ stacking training mechanisms and ensemble learning, they may not be directly comparable with our approach. The results in Table \ref{tab:com_baseline} show that MTQ-Net performs better than MOS-SSL in S-MOS and G-MOS prediction, but worse than MOS-SSL in N-MOS prediction. Overall, MTQ-Net outperforms MOS-SSL, which confirms the benefits of using the MPL approach to deploy MTQ-Net. It is worth mentioning that unlike MOS-SSL, which requires training a separate model to predict each assessment score, a single MTQ-Net model can simultaneously predict S-MOS, N-MOS, and G-MOS scores given an audio waveform as input.

\begin{table}[t]
\caption{LCC, SRCC, and MSE results of MTQ-Net and MOS-SSL.}
\footnotesize
\label{tab:com_baseline}
\begin{center}
 \begin{tabular}{c||c||c||c} 
 \hline
 \hline
 \textbf{Systems} &\textbf{LCC} & \textbf{SRCC} & \textbf{MSE}  \\ [0.5ex] \cline{2-3}
 \hline\hline
  \multicolumn{4}{c} {S-MOS Score Prediction} \\
 \hline
MOS-SSL &0.904&0.903&0.056\\ \hline
MTQ-Net &\textbf{0.912}&\textbf{0.903}&\textbf{0.043}\\ \hline
 \hline
  \multicolumn{4}{c} {N-MOS Score Prediction} \\
 \hline
MOS-SSL &\textbf{0.770}&\textbf{0.811}&\textbf{0.093}\\ \hline
MTQ-Net &0.739&0.789&0.108\\ \hline 

 \hline
  \multicolumn{4}{c} {G-MOS Score Prediction} \\
 \hline
MOS-SSL &0.849&0.852&0.052\\ \hline
MTQ-Net &\textbf{0.881}&\textbf{0.882}&\textbf{0.039}\\ \hline
\end{tabular}
\end{center}
\end{table}

\section{Conclusions}
In this paper, we have proposed the MPL approach to achieve robust prediction capabilities of speech assessment models. MPL consists of obtaining pseudo-label scores from a well-trained speech assessment model and performing multi-task learning to deploy the target speech assessment model. 
In this study, the well-trained model is the MOSA-Net model for predicting PESQ, STOI, and SDI scores, while the target model is the MTQ-Net model for predicting S-MOS, N-MOS, and G-MOS scores.
Our experiments lead to the following conclusions. First, we confirm the advantages of MPL over knowledge transfer and training from scratch approaches, enabling the deployed MTQ-Net model to achieve better prediction capabilities. Second, we validate the benefits of using Huber loss, which combines the strengths of mean absolute error (MAE) and mean squared error (MSE), to achieve improved prediction performance. Third, MTQ-Net trained with the MPL approach can achieve overall higher prediction performance compared to the strong SSL-based model MOS-SSL. In the future, we will investigate potential integration of MTQ-Net with speech processing applications.

\bibliographystyle{IEEEbib}
\bibliography{refs}

\end{document}